\documentstyle[12pt,preprint,aps,prb,eqsecnum]{revtex}
\def\nm{\nonumber}
  
\def\beqa{\begin{eqnarray}}
\def\beq{\begin{equation}}  
\def\eeqa{\end{eqnarray}}
\def\eeq{\end{equation}}

\def\L{{\cal{L}}}   
\def\lab{\label}   
\def\pa{\partial}
\def\l{\Lambda}
\def\O{\Omega}
\def\wO{\widehat{\Omega}}

\begin{document}

\begin{titlepage}
\thispagestyle{plain}
\pagenumbering{arabic}
\vspace*{-1.9cm}
\vspace{1.0cm}
%%%%%%%%%%%%%%%%%%%%%%%%%%%%%%%%%%%%%%%%%%%%%%%%%%%%%
\begin{center}
{\Large \bf Picard-Fuchs Equations and Whitham Hierarchy in }
\end{center}
\begin{center}
{\Large \bf $N=2$ Supersymmetric SU($r+1$) Yang-Mills Theory}
\end{center}
\vspace{-7.0mm}
%%%%%%%%%%%%%%%%%%%%%%%%%%%%%%%%%%%%%%%%%%%%%%%%%%%%%%%%%
\lineskip .80em
\vskip 4em
\normalsize
\begin{center}
{\large Y\H uji Ohta}
\end{center}
\vskip 1.5em
\begin{center}
{\em Research Institute for Mathematical Sciences }
\end{center}
\vspace{-10.5mm}
\begin{center}
{\em Kyoto University}
\end{center}
\vspace{-10.5mm}
\begin{center}
{\em Sakyoku, Kyoto 606, Japan.}
\end{center}
%%%%%%%%%%%%%%%%%%%%%%%%%%%%%%%%%%%%%%%%%%%%%%%%%%%%%%%%%%
\vspace{1.0cm}
\begin{abstract}
In general, Whitham dynamics involves infinitely many parameters 
called Whitham times, but in the context of $N=2$ supersymmetric Yang-Mills 
theory it can be regarded as a finite system by restricting the number of 
Whitham times appropriately. For example, in the case of SU($r+1$) gauge 
theory without hypermultiplets, there are $r$ Whitham times and they 
play an essential role in the theory. 
In this situation, the generating 
meromorphic 1-form of the Whitham hierarchy 
on Seiberg-Witten curve is represented by a finite linear 
combination of meromorphic 1-forms associated with these Whitham times, 
but it turns out that there are various differential relations among these 
differentials. Since these relations can be written only in terms of 
the Seiberg-Witten 1-form, their consistency conditions are found to 
give the Picard-Fuchs equations for the Seiberg-Witten periods. 
\end{abstract}
%%%%%%%%%%%%%%%%%%%%%%%%%%%%%%%%%%%%%%%%%%%%%%%%%%%%%%%%%%%%%%%%%
\end{titlepage}

%\pagestyle{myheadings}
%\markboth{Picard-Fuchs equations and Whitham hierarchy}
%{Picard-Fuchs equations and Whitham hierarchy}

%%%%%%%%%%%%%%%%%%%%%%%%%%%%%%%%%%%%%%%%%%%%%%%%%%%%%%%%%%%%%%%%%%%%%
\begin{center}
\section{Introduction}
\end{center}

Thanks to the study of electro-magnetic duality initiated by Seiberg and 
Witten, \cite{SW} the prepotential of the low energy effective action of 
$N=2$ supersymmetric Yang-Mills theory was turned out to be viewed as a 
function on a complex projective space having singularities when 
the masses of charged particles vanish. This complex projective 
space can be identified with the moduli space of a Riemann surface 
determined by several physical requirements, thus the effective 
theory can be considered to be controlled by the geometry of moduli space 
of a Riemann 
surface. \cite{KLYT,AF,HO,MN,DS1,DS2,Han,BL,APS,AAM,LPG,AAG,MW,LW,EY} 
According to this observation, 
since the effective coupling constants of the theory is interpreted as 
the period matrix of a Riemann surface, determining the period matrix from 
calculation of periods becomes equivalent to 
evaluate effective coupling constants. It is interesting 
that the instanton contributions to prepotential \cite{Sei} 
can be obtained from the evaluation of periods and the prepotentials 
obtained in this way \cite{KLT,IY,O,Matone,EFT,EF,Ito,Ghe,IS} 
are known to be consistent to the instanton 
calculus. \cite{IS,FP,IS3,Slat,DKM,HS,AHSW,Y} 
In these studies, the method based on Picard-Fuchs equations 
\cite{KLT,IY,O,Matone,EFT,EF,Ito,Ghe,Ryang,GSA,IMNS,Ali,O3} played a 
crucial role. 

However, on the one hand, the theory of prepotential often shows 
unexpected aspects behind the effective theory. For example, 
it is known that the Seiberg-Witten solutions can be understood in the 
framework of Whitham theory. \cite{GKMMM,NT} Gorsky {\em et al.} \cite{GMMM} 
noticed that the Whitham dynamics in $N=2$ Yang-Mills theory could 
be written essentially by only finite number of Whitham times and 
found that the second-order derivatives of prepotential over the 
Whitham times could be represented by an elliptic function associated with 
Seiberg-Witten curve. 

However, we can further learn more aspects of Whitham hierarchy 
in gauge theory from the basic idea of Gorsky {\em et al.}. \cite{GMMM} 
For instance, note that since the number of time variables of the hierarchy 
is restricted to be finite the generating meromorphic 
1-form of the Whitham hierarchy is represented by a finite linear 
combination of meromorphic 1-forms associated with these Whitham times. 
Then we can expect that there must be closed differential relations among 
these meromorphic differentials associated with Whitham times. 
In fact, a detailed study supports this 
observation and the aim of the paper is to show the consequence of 
these relations, especially, a connection to Picard-Fuchs equations 
for the Seiberg-Witten periods. 

The paper is organized as follows. In Sec. II, we briefly summarize the 
Whitham dynamics in SU($r+1$) gauge theory. In addition, following to 
Gorsky {\em et al.}, \cite{GMMM} we consider the 
situation that the number of Whitham times is finite. 
Since the meromorphic 1-forms on Seiberg-Witten curve consisting of the 
Whitham hierarchy must be always represented by simply a linear combination 
of Abelian differentials, we can expect the existence of differential 
relations among these meromorphic 1-forms. In Sec. III, it is shown that 
such relations can be in fact found and as a result 
Picard-Fuchs equations for the Seiberg-Witten periods are available 
from this view point. It should be noted 
that the generating meromorphic differential of the Whitham hierarchy can 
be written in terms of the Seiberg-Witten 1-form. This indicates that 
it is sufficient to consider only the Seiberg-Witten periods in order to 
calculate the periods of the Whitham hierarchy. In Sec. IV, it is 
shown that the SU(3) Picard-Fuchs equations for the Seiberg-Witten periods 
can be obtained from the Picard-Fuchs equations with Whitham times 
for the periods of the Whitham hierarchy by considering the specialization 
condition to Seiberg-Witten model. Sec. V is a brief summary. 

%%%%%%%%%%%%%%%%%%%%%%%%%%%%%%%%%%%%%%%%%%%%%%%%%%%%%%%%%%%%%%%%%%%%%
\begin{center}
\section{Whitham hierarchy in gauge theory}
\end{center}

In this section, we briefly sketch the relation between Seiberg-Witten 
solution and Whitham dynamics in the context of 
$N=2$ supersymmetric Yang-Mills theory. \cite{GKMMM,NT,GMMM}

%%%%%%%%%%%%%%%%%%%%%%%%%%%%%%%%%%%%%%%%%%%%%%%%%%%%%%%%%%%%%%%%%%%%%%
\begin{center}
\subsection{Seiberg-Witten solution}
\end{center}

To begin with, let us recall that the Seiberg-Witten curve in SU($r+1$) 
gauge theory without matter hypermultiplets \cite{KLYT,AF,HO} 
is given by the characteristic equation 
	\beq
	\det [x-L(\omega )]=0
	\lab{lax}
	\eeq
of the Lax operator $L(\omega )$ for Toda chain with $r+1$ 
sites, \cite{GKMMM} where $x$ is the eigenvalue of $L(\omega )$ and 
$\omega$ is the spectral parameter. (\ref{lax}) can be rewritten in the 
form of spectral curve 
	\beq
	P(x)=\l_{\mbox{\scriptsize SU($r+1$)}}^{r+1}
	\left( \omega +\frac{1}{\omega}\right)
	\lab{Luke-Skywalker}
	,\eeq
where $\l_{\mbox{\scriptsize SU($r+1$)}}$ is the dynamical mass parameter 
and  
	\beq
	P(x):=x^{r+1} -\sum_{i=2}^{r+1}u_i x^{r+1-i}
	\eeq
represents the simple singularity of type $A_r$ with moduli $u_i$. This 
spectral curve (\ref{Luke-Skywalker}) can be further rewritten in the 
familiar hyperelliptic form \cite{KLYT,AF,HO}  
	\beq
	y^2 =P^2 -4\l^2
	,\lab{SWc}
	\eeq
where $\l :=\l_{\mbox{\scriptsize SU($r+1$)}}^{r+1}$ and we have 
introduced  
	\beq
	y :=\l_{\mbox{\scriptsize SU($r+1$)}}^{r+1}
	\left( \omega -\frac{1}{\omega}\right)
	.\eeq
Note that the hyperelliptic curve (\ref{SWc}) is a Riemann surface of 
genus $r$. 

For a study of Riemann surface, it is often useful to consider 
the periods of Abelian differentials over the 1-cycles on the surface. 
In the case at hand, we can take $2r$ 1-cycles 
$(A_i ,B_i )$ $(i=1,\cdots,r)$ on (\ref{SWc}) as a canonical basis ($B_i$ 
are symplectic duals of $A_i$), which can be expressed by using the 
branching points of (\ref{SWc}). 

On the other hand, in order to interpret the components of period 
matrix constructed from periods of Abelian differentials as the 
effective coupling constants, the combination of Abelian differentials must 
be fixed uniquely up to total derivatives. In addition, in general, there 
are three kinds of Abelian differentials on a Riemann surface, but that of 
the third kind is not required here because we are considering a pure gauge 
theory. Therefore, the expected meromorphic differential 1-form is 
expressed by the Abelian differentials of the first and second kinds, and 
the one satisfying these requirements is called Seiberg-Witten 
differential $dS_{\mbox{\scriptsize SW}}$, given by 
	\beq
	dS_{\mbox{\scriptsize SW}} :=x\frac{d\omega}{\omega}
	=x\frac{\pa_x P}{y}dx
	\lab{26}
	,\eeq
where we have ignored the numerical normalization for simplicity, 
and then the Seiberg-Witten periods are given by the 
loop integrals over the canonical cycles 
	\beq
	a_i :=\oint_{A_i}dS_{\mbox{\scriptsize SW}}
	,\ a_{D_i}:=
	\oint_{B_i}dS_{\mbox{\scriptsize SW}}
	.\eeq 
Note that $dS_{\mbox{\scriptsize SW}}$ can be viewed as the canonical 1-form 
of the integrable system. In this way, we can see the relation between 
Seiberg-Witten solution and integrable system. 

%%%%%%%%%%%%%%%%%%%%%%%%%%%%%%%%%%%%%%%%%%%%%%%%%%%%%%%%%%%%%%%%%%%%%
\begin{center}
\subsection{Whitham hierarchy}
\end{center}

We have seen that the Seiberg-Witten solution has a connection to integrable 
system, but it can be also viewed as a part of Whitham theory of solitons 
on a Riemann surface. 

To see this, let us recall that in general Whitham theory consists of the 
following three ingredients: \cite{Kri} 
	\begin{itemize}
	\item 
	Riemann surface of genus $g$. 
	\item 
	Punctures on the surface. 
	\item 
	Existence of local coordinates near the punctures.
	\end{itemize}

Gorsky {\em et al.} \cite{GMMM} noticed that the meromorphic differentials 
of the second kind $d\O_n$ of $(n+1)$-th order punctures $(n>0)$ on a 
Riemann 
surface was defined up to a linear combination of $g$ holomorphic 
differentials $d\omega_i$ and considered how to fix this combination by 
taking two basic requirements. The first one was to require 
	\beq
	\oint_{A_i}d\O_n =0
	\eeq
and the second one was to introduce new meromorphic differentials $d\wO_n$ 
which enjoy the property that their differentiations over the moduli 
coincide with holomorphic differentials. 

According to their result, \cite{GMMM} the differential 
	\beq
	dS :=\sum_{n=1}^{\infty}T_n d\wO_n =\sum_{i=1}^{g}\alpha_i 
	d\omega_i +\sum_{n=1}^{\infty}T_n d\O_n
	\lab{29}
	\eeq
with infinitely many parameters $T_n$ called Whitham times is found to be 
the expected solution which is suitable for applications to gauge theory. 
For this new meromorphic differential $dS$, the 
periods 
	\beq 
	\alpha_i :=\oint_{A_i}dS ,\ \alpha_{D_i}:=\oint_{B_i} dS
	\lab{211}
	\eeq
can be defined in a natural way. 

Next, in order to make a contact with Seiberg-Witten solution, 
Gorsky {\em et al.} \cite{GMMM} regarded the Riemann surface used here as 
the Seiberg-Witten hyperelliptic curve (\ref{SWc}). 

In such a situation, they found that the Whitham hierarchy could be 
actually written by only first $r$ time variables and gave an explicit 
expression of $dS$. In particular, in the case of SU($r+1$) gauge theory, 
$n$ is restricted to $n<r+1$. Namely, in this situation, the periods 
(\ref{211}) reduce to  
	\beq
	\alpha_i =\sum_{n=1}^{r}T_n \oint_{A_i}d\wO_n 
	,\ \alpha_{D_i} =\sum_{n=1}^{r}T_n \oint_{B_i}d\wO_n 
	\lab{Yoda}
	\eeq
and $d\wO_n$ are given by 
	\beq
	d\wO_n =R_n \frac{d\omega}{\omega},\ R_n :=P_{+}^{n/(r+1)}
	.\lab{33}
	\eeq
In this expression, $P_{+}^{n/(r+1)}$ means the non-negative terms 
in the expansion of $P^{n/(r+1)}$ for a large $x$, and in general, 
$P^{n/(r+1)}$ in SU($r+1$) gauge theory is easily found to give 
	\beq
	P^{n/(r+1)}  =x^n -\frac{n}{r+1}u_2 x^{n-2}-\frac{n}{r+1}u_3 x^{n-3}-
	\frac{n}{r+1}\left[u_4 +\frac{u_{2}^2}{2}\left(1-\frac{n}{r+1}
	\right)\right]x^{n-4}-\cdots 
	.\eeq

Note that the periods are now represented by a finite linear 
combination of $d\wO_n$ because we are considering only for $n<r+1$ case. 
In addition, from (\ref{33}), it is immediate to see that the 
Seiberg-Witten solution is recovered at the point 
	\beq
	(T_1 ,T_2 , T_3 , \cdots ,T_r )=
	(1,0,0,\cdots ,0)
	\lab{Mr.Spok}
	.\eeq
In fact, we find $d\wO_1 =dS_{\mbox{\scriptsize SW}}$. Of course, 
in this case, we have $dS=dS_{\mbox{\scriptsize SW}}$. 

%%%%%%%%%%%%%%%%%%%%%%%%%%%%%%%%%%%%%%%%%%%%%%%%%%%%%%%%%%%%%%%%%%%%%
\begin{center}
\section{Picard-Fuchs structure behind Whitham hierarchy}
\end{center}

%%%%%%%%%%%%%%%%%%%%%%%%%%%%%%%%%%%%%%%%%%%%%%%%%%%%%%%%%%%%%%%%%%%%%
\begin{center}
\subsection{Relations among meromorphic differentials}
\end{center}

We have seen that $dS$ is represented by a linear combination 
of $d\wO_n$ and also seen that $d\wO_1 =dS_{\mbox{\scriptsize SW}}$. 
Then, are $d\wO_n$ for $n \neq 1$ related to 
$dS_{\mbox{\scriptsize SW}}$? If we can find any relation among them, 
the role of the Seiberg-Witten solution in the Whitham dynamics will be 
clarified. 

To find an answer to this question, let us notice that any 
meromorphic differential on a Riemann surface must be always 
written in terms of the basis of Abelian differentials on the 
surface. Of course, this must be true also for $d\wO_n$ for 
all $n$. Therefore, if we consider a differentiation of $d\wO_n$ 
over moduli, it will be ultimately represented by a linear combination of 
various $d\wO_n$ and their derivatives. 
However, actually, in the case of Seiberg-Witten Riemann surface, 
we can show that the derivatives of 
$d\wO_n$ for $n>1$ are obtained from the Seiberg-Witten differential 
$d\wO_1$. Thus as the result, we can conclude that 
$dS$ is generated from $d\wO_1$ and accordingly the periods of $dS$ can be 
directly determined through the Seiberg-Witten periods themselves. 

To see this more concretely, let us consider the case of $d\wO_2$ as an 
example. Since the differentiations of $d\wO_2$ over moduli are  
	\beq
	\frac{\pa d\wO_2}{\pa u_i}=\frac{dx}{y}
	\left[ -2\delta_{2,i}x^{r} +\frac{\delta_{2,i}}{r+1}\sum_{j=2}^{r+1}
	(r+1-j)u_j x^{r-j} +2x^{r+2-i}\right]
	,\lab{31}
	\eeq
where $\delta_{i,j}$ are the Kronecker's delta symbols, and those 
for $d\wO_1$ are 
	\beq
	\frac{\pa d\wO_1}{\pa u_i}=\frac{x^{r+1-i}}{y}dx
	,\lab{32}
	\eeq
it is easy to see that 
	\beq
	\frac{\pa d\wO_2}{\pa u_2}=\frac{2}{r+1}\sum_{i=2}^{r+1}(r+1-i)
	u_i \frac{\pa d\wO_1}{\pa u_{i+1}},\ 
	\frac{\pa d\wO_2}{\pa u_i}=2\frac{\pa d\wO_1}{\pa u_{i-1}} \ \ \ 
	(i\neq 2)
	.\lab{C3PO} 
	\eeq
Note that in the derivation of (\ref{31}) and (\ref{32}) we have used the 
general formulae
	\beq
	\frac{\pa d\wO_n}{\pa u_i}=\frac{dx}{y}\left[\pa_{u_i}R_n \cdot 
	\pa_x P -\pa_x R_n \cdot \pa_{u_i}P\right]+
	d\left(\frac{R_n \pa_{u_i}P}{y}\right)
	.\eeq

In a similar way, we can obtain differential relations between 
$d\wO_n$ for $n>1$ and $d\wO_1$, but we omit the derivations for them 
and show only the result for $n=3$ and $4$ cases here. 

For $d\wO_3$:
	\beqa
	& &\frac{\pa d\wO_3}{\pa u_2}=-\frac{3}{r+1}\left[
	u_2 \frac{\pa d\wO_1}{\pa u_2} -\sum_{i=2}^{r+1}(r+1-i)u_i 
	\frac{\pa d\wO_1}{\pa u_i }\right],\nm\\
	& &\frac{\pa d\wO_3}{\pa u_3}=-\frac{3}{r+1}\left[
	u_2 \frac{\pa d\wO_1}{\pa u_3} -\sum_{i=2}^{r+1}(r+1-i)u_i 
	\frac{\pa d\wO_1}{\pa u_{i+1}}\right],\nm\\
	& &\frac{\pa d\wO_3}{\pa u_i}=3\left[ \frac{\pa d\wO_1}{\pa u_{i-2}}
	-\frac{u_2}{r+1}\frac{\pa d\wO_1}{\pa u_i}\right] \ \ \ 
	(i\neq 2,3)
	.\lab{35}
	\eeqa

For $d\wO_4$:
	\beqa
	\frac{\pa d\wO_4}{\pa u_2}&=&-\frac{4}{r+1}\Biggl[
	u_3 \frac{\pa d\wO_1}{\pa u_2}-
	\sum_{i=3}^{r+1}(r+1-i)u_i \frac{\pa d\wO_1}{\pa u_{i-1}}
	-\frac{r-3}{r+1}u_2 \sum_{i=2}^{r+1}(r+1-i)u_i \frac{\pa d\wO_1}
	{\pa u_{i+1}}\Biggr],\nm\\
	\frac{\pa d\wO_4}{\pa u_3}&=&-\frac{4}{r+1}\left[ 2u_2 \frac{\pa d
	\wO_1}{\pa u_2}+u_3 \frac{\pa d\wO_1}{\pa u_3}-\sum_{i=2}^{r+1}
	(r+1-i)u_i \frac{\pa d\wO_1}{\pa u_i}\right],\nm\\ 
	\frac{\pa d\wO_4}{\pa u_4}&=&-\frac{4}{r+1}\left[ 2u_2 \frac{\pa d
	\wO_1}{\pa u_3}+u_3 \frac{\pa d\wO_1}{\pa u_4}-\sum_{i=2}^{r+1}
	(r+1-i)u_i \frac{\pa d\wO_1}{\pa u_{i+1}}\right],\nm\\
	\frac{\pa d\wO_4}{\pa u_i}&=&4\left[ \frac{\pa d
	\wO_1}{\pa u_{i-3}}-\frac{2u_2}{r+1}\frac{\pa d\wO_1}{\pa u_{i-1}}-
	\frac{u_3}{r+1}\frac{\pa d\wO_1}{\pa u_i}\right] \ \ (i\neq 2,3,4 )
	\lab{36}
	.\eeqa

%%%%%%%%%%%%%%%%%%%%%%%%%%%%%%%%%%%%%%%%%%%%%%%%%%%%%%%%%%%%%%%%%%%%%
\begin{center}
\subsection{Picard-Fuchs equations from Whitham hierarchy}
\end{center}

If the derivatives of $d\wO_n$ over moduli for $n>1$ are eliminated from 
the relations (\ref{C3PO}), (\ref{35}) and (\ref{36}) by using 
differentiations, the equations satisfied by $d\wO_1$ will be obtained. 
Furthermore, since $d\wO_1 =dS_{\mbox{\scriptsize SW}}$, 
we can identify such equations as Picard-Fuchs equations for 
Seiberg-Witten periods. 

To see this, it is enough to consider the cross derivatives of $d\wO_n$. 
For example, for $d\wO_2$, from $[\pa_2 \pa_i -\pa_i \pa_2]d\wO_2 =0$, 
where $\pa_i :=\pa /\pa u_i$, we get 
	\beq
	\left[ (r+1)\pa_2 \pa_{i-1} -(r+1-i)\pa_{i+1}-\sum_{j=2}^{r+1}
	(r+1-j) u_j \pa_i \pa_{j+1} \right]d\wO_1 =0 \ \ (i\neq 2) 
	\lab{R2D2}
	,\eeq
which are the Picard-Fuchs equations obtained by several 
authors. \cite{IS,Ali} For other $d\wO_n$, we can construct similar 
equations and, in fact, we can obtain the ``hierarchy'' of Picard-Fuchs 
equations as follows: 
	\beqa
	& &\left[\, \sum_{i=2}^{r+1}(r+1-i)u_i (\pa_3 \pa_i -\pa_2 
	\pa_{i+1})\right]d\wO_1 =0,\nm\\
	& &\left[ (r+1-i)\pa_{i+1} -(r+1)\pa_3 \pa_{i-2}+
	\sum_{j=2}^{r+1}(r+1-j)u_j \pa_i \pa_{j+1}\right]d\wO_1 =0\ \ 
	(i\neq 2,3),\nm\\ 
	& &\left[(r+1)\pa_2 \pa_{i-2}-(r+2-i)\pa_i -
	\sum_{j=2}^{r+1}(r+1-j)u_j \pa_i \pa_j \right]
	d\wO_1 =0 \ \ (i\neq 2,3),\nm\\
	& &\Biggl[ 2(r+1)u_2 \pa_{2}^2 +(r-3)(r-2)u_2 \pa_4 +
	(r+1)\sum_{i=3}^{r+1}(r+1-i)u_i \pa_{i-1}\pa_3 \nm\\
	& &\qquad\quad +(r-3)u_2 \sum_{i=2}^{r+1}(r+1-i)u_i \pa_{i+1}\pa_3 
	-(r+1)\sum_{i=2}^{r+1}(r+1-i)u_i \pa_i \pa_2 \Biggr]d\wO_1 =0,\nm\\
	& &\Biggl[ 2(r+1)u_2 \pa_2 \pa_3 +(r-3)^2 u_2 \pa_5 +(r+1)
	\sum_{i=3}^{r+1}(r+1-i)u_i \pa_{i-1}\pa_4 \nm\\
	& &\qquad\quad -(r+1)\sum_{i=2}^{r+1}(r+1-i)u_i \pa_{i+1}\pa_2 +
	(r-3)u_2 \sum_{i=2}^{r+1}(r+1 -i)u_i \pa_{i+1}\pa_4
	\Biggr]d\wO_1 =0,\nm\\
	& &\Biggl[ (r+1)^2 \pa_2 \pa_{i-3}-2(r+1)u_2 \pa_{i-1}\pa_2 
	-(r+1)(r+3-i)\pa_{i-1}\nm\\
	& &\qquad\quad -(r-3)(r+1-i)u_2 \pa_{i+1}
	-(r+1)\sum_{j=3}^{r+1}(r+1-j)u_j \pa_i \pa_{j-1}\nm\\
	& &\qquad\quad 
	-(r-3)u_2 \sum_{j=2}^{r+1}(r+1-j)u_j \pa_i \pa_{j+1}\Biggr]
	d\wO_1 =0 \ \ (i\neq 2,3,4),\nm\\
	& &\left[ 2u_2 (\pa_2 \pa_4 -\pa_{3}^2 )-
	\sum_{i=2}^{r+1}(r+1-i)u_i (\pa_4 \pa_i -\pa_3 \pa_{i+1})\right]
	d\wO_1 =0,\nm\\
	& &\Biggl[ 2u_2 (\pa_3 \pa_i -\pa_4 \pa_{i-1})+(r+1)
	\pa_4 \pa_{i-3}-(r+1-i)\pa_{i+1}\nm\\
	& &\qquad\quad -
	\sum_{j=2}^{r+1}(r+1-j)u_j \pa_i \pa_{j+1}
	\Biggr]d\wO_1 =0 \ \ (i\neq 2,3,4),\nm\\
	& &\Biggl[ 2u_2 (\pa_i \pa_2 -\pa_3 \pa_{i-1})+(r+1)\pa_3 \pa_{i-3}
	-(r+2-i)\pa_i \nm\\
	& &\qquad\quad -\sum_{j=2}^{r+1}(r+1-j)u_j 
	\pa_i \pa_j \Biggr]d\wO_1 =0 \ \ (i\neq 2,3,4)
	.\lab{310}
	\eeqa

Note that the equations in (\ref{310}) are all second-order equations and 
in some cases we can simplify them by using 
$(\pa_i \pa_j -\pa_p \pa_q )d\wO_1 =0$, where $i+j =p+q$. \cite{IS,Ali}

%%%%%%%%%%%%%%%%%%%%%%%%%%%%%%%%%%%%%%%%%%%%%%%%%%%%%%%%%%%%%%%%%%%%%
\begin{center}
\subsection{Picard-Fuchs equations as a complete system}
\end{center}

Of course, as a complete Picard-Fuchs system, it is not necessary to 
consider all equations in (\ref{R2D2}) and (\ref{310}). In general, since 
there are $r$ moduli parameters in the SU($r+1$) gauge theory, it is 
sufficient to extract at least $r$ independent equations from them. 

To see this, let us notice the equations in (\ref{R2D2}). 
Since the number of the equations is $r-1$, one more equation is 
necessary. However, we can not obtain the expected equation from (\ref{310}) 
because the equations presented there do not have the instanton corrections. 
If the instanton correction terms are not included in any one of 
Picard-Fuchs equations, the prepotential obtained from them will 
not show the instanton corrections precisely. Therefore, 
we require that the remaining one must include instanton terms. 

Actually, such equation was recognized by Ito and Yang \cite{IYY} 
as the scaling relation. There, the Picard-Fuchs system was 
realized by two kinds of equations, one of which is Gauss-Manin system and 
the other is the scaling relation. Since the Gauss-Manin system does not 
involve instanton corrections, the situation looks like our's. Therefore, 
also for our case, the scaling relation may be used as the 
remaining Picard-Fuchs equation. 

For this, let us consider the Eulerian operator 
	\beq
	{\cal{E}}:=\sum_{i=2}^{r+1}iu_i \pa_i +(r+1)\l \pa_{\l}
	\lab{Eu}
	,\eeq
which acts as 
	\beq
	{\cal{E}}d\wO_n =nd\wO_n 
	\lab{Eu2}
	\eeq
for all $n>0$. (\ref{Eu2}) indicates that the degree of $d\wO_n$ is $n$. 
Realizing (\ref{Eu2}) as an equation only in terms of 
moduli derivatives can be easily accomplished by considering the squaring 
equation $({\cal{E}}-n)^2 d\wO_n =0$. \cite{Ali,IYY} 

In this way, we can associate $r$ independent Picard-Fuchs equations 
for $d\wO_1$. 

%%%%%%%%%%%%%%%%%%%%%%%%%%%%%%%%%%%%%%%%%%%%%%%%%%%%%%%%%%%%%%%%%%%%%
\begin{center}
\section{Picard-Fuchs equations with Whitham times}
\end{center}

%%%%%%%%%%%%%%%%%%%%%%%%%%%%%%%%%%%%%%%%%%%%%%%%%%%%%%%%%%%%%%%%%%%%%
\begin{center}
\subsection{The SU(3) Picard-Fuchs equations}
\end{center}

Next, let us consider Picard-Fuchs equations for the periods 
$(\alpha_i ,\alpha_{D_i})$ of Whitham hierarchy. In the case of $r=1$, 
the resulting Picard-Fuchs equation takes the same form with the usual 
one \cite{KLT} up to rescaling of $T_1$. For this reason, we do not discuss 
this case, and instead, let us consider $r=2$ case in order to find a 
non-trivial example of Picard-Fuchs equations with Whitham times. 

In this case, the Picard-Fuchs equations with the Whitham times are 
found to be in the form 
	\beq
	\L_j (\alpha_i ,\alpha_{D_i}) =0
	,\eeq 
where
	\beqa
	\L_1&=&
	\frac{54u^2 vT_1 T_2 -(108\l^2 -4u^3 -27v^2 )(uT_{2}^2 -
	3T_{1}^2 )}{u(-3T_{1}^2 +4uT_{2}^2 )}\pa_{u}^2 \nm\\
	& &-\frac{
	3[ (-108\l^2 +4u^3 +27v^2 )T_1 T_2 +8uv(uT_{2}^2 -3T_{1}^2 )]}{
	2(3T_{1}^2 -4uT_{2}^2 )}\pa_u \pa_v \nm\\
	& &+\frac{T_2 [8(108\l^2 -4u^3 -27v^2 )(-3T_{1}^2 +uT_{2}^2 )
	T_2 -9uvT_1 (15T_{1}^2 +28uT_{2}^2 )]}{2u(3T_{1}^2 -4uT_{2}^2 )^2}
	\pa_u \nm\\
	& &+\frac{2(108\l^2 -4u^3 -27v^2 )(3T_{1}^2 -uT_{2}^2 )T_1 T_2 
	+3uv(9T_{1}^4 +27uT_{1}^2 T_{2}^2 -4u^2 T_{2}^4)}{
	u(3T_{1}^2 -4uT_{2}^2 )^2}\pa_v +1,\nm\\
	\L_2&=&\frac{-54u^2 vT_1 T_2 +(108\l^2 -4u^3 -27v^2 )(uT_{2}^2 -
	3T_{1}^2 )}{3(3T_{1}^2 -4uT_{2}^2 )}\pa_{v}^2 \nm\\
	& &-\frac{3[8uv(uT_{2}^2-3T_{1}^2)+(-108\l^2 +4u^3 +27v^2 )
	T_1 T_2 ]}{2(3T_{1}^2 -4uT_{2}^2 )}\pa_u \pa_v 
	-\frac{45vT_1 T_2}{2(3T_{1}^2 -4uT_{2}^2 )}\pa_u \nm\\
	& &+\frac{3v(3T_{1}^2 +uT_{2}^2 )}{3T_{1}^2 -4uT_{2}^2}\pa_v +1
	.\lab{pff}
	\eeqa

Though the derivation of Picard-Fuchs equations for other higher $r$ is 
straightforward, the result is turned out to be too lengthy and 
complicated, so we do not consider these cases in this paper. 

%%%%%%%%%%%%%%%%%%%%%%%%%%%%%%%%%%%%%%%%%%%%%%%%%%%%%%%%%%%%%%%%%%%%%
\begin{center}
\subsection{Specializations of SU(3) Picard-Fuchs equations}
\end{center}

It may be instructive to see specializations of (\ref{pff}). 
With the help of (\ref{Mr.Spok}), 
it is straightforward to make sure that the equations in (\ref{pff}) yield 
the usual SU(3) Picard-Fuchs equations \cite{KLT} $L_j (a_i ,a_{D_i})=0$ 
for the Seiberg-Witten periods, which can be identified with 
Appell's $F_4$ system \cite{Er,Tak,Kato,Kan} 
	\beqa
	& &\L_1 \rightarrow L_1 :=(4u^3 +27v^2 -108\l^2 )\pa_{u}^2 
	+12u^2 v \pa_u \pa_v +3uv\pa_v +u,\nm\\
	& &\L_2 \rightarrow L_2 :=(4u^3 +27v^2 -108\l^2 )\pa_{v}^2 
	+36uv\pa_u \pa_v +9v\pa_v +3
	\lab{L}
	.\eeqa 
Note that the consistency condition of (\ref{L}) leads to 
$[u\pa_{v}^2 -3\pa_{u}^2] (a_i ,a_{D_i})=0$, 
which coincides with (\ref{R2D2}) for $r=2$. 

On the other hand, from (\ref{pff}) with $(T_1 ,T_2 )=(0,1)$, 
we can also consider Picard-Fuchs equations 
$\widehat{L}_j (\oint_{A_i}d\wO_2 ,\oint_{B_i}d\wO_2 )=0$ for the periods 
of $d\wO_2$, where 
	\beqa
	\L_1 \rightarrow \widehat{L}_1 &:=&u(4u^3 +27v^2 -108\l^2 )\pa_{u}^2 
	+12u^3 v\pa_u \pa_v -
	(4u^3 +27v^2 -108\l^2 )\pa_u -3u^2 v\pa_v +4u^2 , \nm\\
	\L_2 \rightarrow \widehat{L}_2 &:=&(4u^3 +27v^2 -108\l^2 )\pa_{v}^2 
	+36uv\pa_u \pa_v -9v\pa_v +12
	\lab{ren}
	.\eeqa
From (\ref{ren}), we can obtain a relation like that from (\ref{L}), but 
the same equation is also available from (\ref{C3PO}), provided 
$\pa d\wO_1 /\pa u_i$ are eliminated from (\ref{C3PO}). 

Finally, note that we have 
	\beq
	\widehat{L}_j (\alpha_i ,\alpha_{D_i}) =T_1 
	\widehat{L}_j\left(\oint_{A_i}d\wO_1 ,\oint_{B_i}d\wO_1 \right)
	,\ L_j (\alpha_i ,\alpha_{D_i}) =T_2 
	L_j \left(\oint_{A_i}d\wO_2 ,\oint_{B_i}d\wO_2 \right)
	\eeq
from (\ref{29}), (\ref{L}) and (\ref{ren}). 

%%%%%%%%%%%%%%%%%%%%%%%%%%%%%%%%%%%%%%%%%%%%%%%%%%%%%%%%%%%%%%%%%%%%%
\begin{center}
\section{Summary}
\end{center}

In this paper, we have discussed the SU($r+1$) gauge theory in the 
standpoint of Whitham dynamics and realized $r-1$ 
Picard-Fuchs equations for Seiberg-Witten periods as consistency 
equations among meromorphic differentials associated with Whitham times. 
In addition, we have used the scaling relation as the remaining independent 
equation in order to include the instanton corrections. Though the 
generalization to other cases except SU($r+1$) group is straightforward, 
the case of exceptional gauge groups would be interesting because there are 
two types of Seiberg-Witten curves in these gauge 
theories. \cite{DS2,AAM,LPG,AAG,MW,LW,EY} In particular, 
it may be interesting to know how the differences of physics expected 
from these two curves \cite{LPG,EY,Ito,Ghe} are reflected in the Whitham 
theory and the Picard-Fuchs structure behind it. 

Of course, our construction of Picard-Fuchs equations may provide 
helpful informations not only for these cases but also 
when we consider the relation among flat coordinates, \cite{IXY,IY2} 
Witten-Dijkgraaf-Verlinde-Verlinde 
equations \cite{Wit,DW,DVV,MMM1,MMM2,IYYY} and Whitham 
hierarchy. \cite{GKMMM,NT,GMMM} We are now planning a discussion 
respect to this point.

%%%%%%%%%%%%%%%%%%%%%%%%%%%%%%%%%%%%%%%%%%%%%%%%%%%%%%%%%%%%%%%%%%%%%
\begin{center}

\end{center}

\end{document}